\documentclass[10pt,draft,notref,notcite]{article}

\usepackage{color,longtable}
\usepackage{amsmath,amssymb,graphicx}
\usepackage{amssymb,graphicx}

\newcommand{\be}{\begin{eqnarray}}
\newcommand{\ee}{\end{eqnarray}}

\newcommand{\Es}[2]{E^{\,\,\,#1}_{(11)#2}}

\usepackage{amsfonts,amsmath}
\usepackage{latexsym}

\def\cL{{\cal{L}}}

\def\cI{{\cal I}}

\def\cO{{\cal O}}

\def\id{{\mathbb{I}}}

\def\cL{{\cal L}}

\def\Es{{\rm E}_{7(7)}}
\def\Ea{{\rm E}_{8(8)}}
\def\En{{\rm E}_9}
\def\Ez{{\rm E}_{10}}

\begin{document}

\begin{center}
{\bf \Large Reflections on Supersymmetry$^*$}\\[7mm]
  Hermann Nicolai\\[4mm]
{\sl  Max-Planck-Institut f\"ur Gravitationsphysik\\
     Albert-Einstein-Institut \\
     M\"uhlenberg 1, D-14476 Potsdam, Germany\\[1mm]
 Email: {\tt nicolai@aei.mpg.de}} \\[17mm]
\begin{minipage}{12cm}\footnotesize
\textbf{Abstract:} Supersymmetry is a theme with many facets
that has dominated much of high energy physics over the 
past decades. In this contribution I present a very personal 
perspective on these developments, which has also been 
shaped in an important way by my interactions with Julius Wess
and Bruno Zumino.
\end{minipage}
\end{center}

\vspace{10cm}

\noindent
*Invited contribution to the second edition of the book
{\it The Supersymmetric World}, ed. by M. Shifman and G. Kane.

\newpage

\section{Introduction}

When Misha Shifman asked me whether I would be willing to
contribute a chapter to this book with some memories 
of my encounters with Julius Wess, I did not hesitate for very long.
For one thing, I am honored to have my name included in such an 
illustrious list of early contributors and pioneers of supersymmetry. Secondly,
because I feel privileged to have been a witness when for a (world-historically 
speaking) fleeting moment the provincial town of Karlsruhe 
became an epicenter of theoretical physics, with Julius Wess and
Bruno Zumino as the central and driving forces behind the development
of supersymmetry. It does not happen often that 
a beginning graduate student is so lucky, as I was, to land in a hot spot 
of new theory developments, especially in a place that never makes
it into the top one hundred of international university rankings.
Of course, there were several other such hot spots, and the 
contributions to this volume vividly and from different angles tell the story 
how the idea of supersymmetry evolved in such diverse places.
For this reason I will not dwell too much on topics that are already extensively 
dealt with in the book, but instead offer my personal perspective, 
covering some aspects that have received less attention 
in the other contributions, especially with regard to $N=8$ supergravity,
as well as some thoughts concerning the current status of supersymmetry
{\em vis-\`a-vis} theoretical high energy physics.

In writing this text I have decided  not to include a list of references.
The reason is simple: there would be either too few, or too many!
At any rate, all references relevant to what I am saying can be easily 
located on the electronic {\tt arXiv}, or alternatively by consulting the 
companion articles in this book.

\section{Early years in Karlsruhe}

Already soon after enrolling at the University of Karlsruhe for my
undergraduate studies in physics it was clear to me that I would want to do my diploma
thesis with Julius Wess who was at the forefront of exciting developments
in theoretical physics. Even as students who knew hardly anything about
the subject, we were keenly aware of this fact! So in 1974 I approached
Julius with the request to become his diploma student. Yet before getting 
accepted I had to overcome a  first hurdle, when he told me that before starting 
in earnest I would have to read the book {\it Introduction to the theory of quantized fields}
by Bogoliubov-Shirkov, the then much used standard 
text on relativistic quantum field theory (QFT). This was clearly the device which
he employed to scare away prospective students who were not $150\%$ willing
to commit. I remained undeterred even though
I can now admit that I never read this formidable book to the end (and
fortunately for me, Julius never checked). This was because 
I realized full well that if I was not accepted, the alternative would have been
to settle for a boring diploma topic, or even (God forbid!) experimental work.

Thanks to the way the institute was run by Julius, life
at the institute was very stimulating in many ways, not
only with regard to physics. This included all kinds of social events 
and gatherings, perhaps most memorably the annual skiing excursions to Oberperfuss
(Austria). The days in Oberperfuss were divided into physics sessions that
took place in an old school building in the mornings and
late afternoons, and ski outings in between.
I was the only institute member who did not know how to ski, 
or even how to stand on skis properly. I still remember how on my first
day I  fell off the lift  immediately on arriving at the top of the slope, 
feeling like the giant beetle in Kafka's novella ``Die Verwandlung".
Nevertheless, after practising for a couple of days I was proudly able to stand and slowly
move on my skis, descending the slope by zigzagging across it  
horizontally (much to the annoyance of people racing down hill), 
which took me a good part of the afternoon.

At the institute in the Physikhochhaus the common room for postdocs and
graduate students bore the label ``Kinder" (children). Indeed we little ones
had hardly any idea what the `adults' ({\em alias} Julius and Bruno Zumino) 
were up to. But we greatly benefitted from the numerous prominent visitors.
Lochlain  O'Raifeartaigh, who discovered the first example of a Wess-Zumino type  
model that exhibits spontaneous breaking of supersymmetry,
spent extended periods of time in Karlsruhe. Another highlight was 
a visit by Rudolf Haag, who started out his seminar with the question ``Wo 
ist Martin Sohnius?". The main purpose of his visit was to work 
with Martin and Jan \L{}opusza\'nski on their now famous classification
of all supersymmetries of the $S$-matrix. Other visitors included 
Andr\'e Martin and John Iliopoulos. John had been scheduled to give a 
supersymmetry seminar about Fayet-Iliopoulos breaking, but switched 
topics to talk about the $c$-quark which had just been discovered, 
and its proper interpretation in terms of the GIM mechanism.
 Pierre Fayet came to present the first version of a supersymmetric
extension of the Standard Model (SM) of particle physics,  inventing a whole 
new vocabulary with neologisms  like `photino', `gluino' 
and the like. Of course, the most prominent visitor was Bruno Zumino 
(which made us joke that there also had to be a superpartner by
the name of `Zumo'). Bruno came to Karlsruhe regularly, but when he 
was around we  rarely saw him, with no idea of what they were doing 
because they usually  locked themselves up, or simply worked at Julius' home. 

In autumn 1975 I finished my diploma thesis on spontaneous breaking of 
global supersymmetry\footnote{We did not publish any of our results because
 L. O'Raifeartaigh scooped us by getting there first! Instead, my first paper (in 1976)
 introduced a non-relativistic quantum mechanical version of supersymmetry, with 
 a supersymmetric spin system built solely from Majorana fermions. This model
 has some similarity with the (supersymmetric) SYK model that became  
 popular only much later.}, which satisfied Julius at least 
 to the extent that he offered me the possibility of staying as a PhD student with him.
As a topic for my doctoral thesis he suggested 
the problem of putting supersymmetry on the lattice. This looked
like a win-win problem for an aspiring PhD student,
because lattice gauge theory was the other freshly hot topic in particle
physics at the time. But, as so often, the devil turned out to hide in the 
details. The idea did not really fly because of one main obstacle: the 
Leibniz rule is no longer valid for lattice derivatives, but nonetheless absolutely needed 
to re-assemble terms into total derivatives, as required for supersymmetry 
invariance of the action. This effectively meant that there was no way
of putting supersymmetry on the lattice without doing violence to it (and 
the problem gets worse if one wants to incorporate gauge symmetry).
My  attempts to get around the difficulty never led to fully satisfactory
answers: for instance, the so-called SLAC derivative
(the Fourier transform of the saw-tooth function $[k^\mu]$) which retains
a rudimentary version of the Leibniz rule, is too non-local 
on the lattice to be of much use. 
Looking for a way out, I turned to constructive QFT, with 
the idea of proving rigorously the existence of Wess-Zumino  model in the framework
of constructive quantum field theory (recall that to this day we have no
example of a fully constructed interacting QFT in four dimensions
satisfying all the Wightman axioms!). At first this looked 
promising because the model required only one divergent (wave-function) 
renormalization, unlike QED or $\phi^4$ theory which need three. Consequently,
one needed to study only a one-dimensional renormalization map, a cutoff dependent 
map between bare and renormalized parameters 
considered in earlier work by R. Schrader. As one of the by-products
this required reformulating supersymmetry in Euclidean signature, a challenge
because the replacement of $SL(2,\mathbb{C})$ by $SU(2)\times SU(2)$ implies
a doubling of the fermions, in apparent conflict with the boson-fermion balance 
of the theory. While others concluded from this that
$N=1$ supersymmetry cannot be Euclideanized,  my proposal 
simply amounted to the statement that the correct Euclideanization 
procedure is the one that yields the proper Euclidean continuation 
of the Minkowskian supersymmetry Ward identities, in accord with
the Osterwalder-Schrader reconstruction theorem. These efforts earned 
me a seminar  invitation to CERN, although I doubt that I was able to convince 
anyone, most importantly Bruno, of the relevance of what I was saying.
In the end my `constructive' attempts were also stymied because the joint UV and  
IR regulators required to rigorously define the functional (path) integral 
break supersymmetry so badly that one could not really exploit the symmetry for its 
full worth. But at least I found a new way to characterize supersymmetry without the 
use of anticommuting variables, just in terms of the bosonic functional measure
that is obtained by integrating out fermions.
 
Meanwhile others at the institute were busy working on all kinds of
different problems spawned by the discovery of supersymmetry, {\em e.g.} 
renormalization of supersymmetric gauge theories (O. Piguet and K. Sibold), 
or possible generalizations of supersymmetry to parasymmetry (N. Dragon).
Considerable effort was spent on searching for off-shell superspace
versions of $N=2$ gauge  theories and supergravity in work involving 
R. Grimm and M. Sohnius, as well as Julius himself. As you can see,
we had a whole new world in front of us to explore!  Of course, Bruno and Julius 
continued their joint work, during the time I was there mostly concentrating
on the superspace of formulation of $N=1$ supergravity, building on the
basic insight that even flat superspace requires torsion. I still recall when
one morning Julius skipped the usual Austrian niceties, by greeting
me simply and directly with the statement ``$\cL = E\,$" -- meaning that the supergravity 
Lagrangian in superspace  is nothing but the supervielbein Berezinian. I 
could see how happy he was with this result,  because 
what could be simpler than that for a gravitational Lagrangian?
A small fly in the ointment was the fact that one still had to take 
into account the superspace constraints, so the seemingly simple result could not be
exploited directly to calculate loop corrections in supergravity.

Although we were all captivated by supersymmetry and in no doubt about
its ultimate relevance, no one in Karlsruhe in those days thought that the available
supersymmetric models could be directly relevant to real physics. Rather, the general
attitude was that essential pieces were still missing from the picture -- just like Yang-Mills
theories are not viable without the crucial extra ingredients discovered
only much later, namely the Brout-Englert-Higgs mechanism, and the subtle
dynamical mechanism of confinement. The main advantage of supersymmetry 
emphasized in those days was not so much the absence of quadratic
divergences (which became the dominant narrative only 
later in connection with the hierarchy problem),
but rather the fact that it overcame the Coleman-Mandula no-go theorem prohibiting 
the fusion of space-time and internal symmetries. The $N=1$
models that became fashionable for phenomenology in  the 80ies precisely 
do {\em not} exploit this feature, which would require extended ($N\geq 2$)
supersymmetry. This is the basic reason for the (to me, esthetically 
unappealing) need to double up the known particle spectrum with unseen
superpartners possessing the same internal quantum numbers.

\section{Beyond simple supersymmetry}

After finishing my PhD in October 1978 and an unsuccessful application
for a CERN fellowship I moved to Heidelberg, where I spent
most of my time learning QCD and perturbative quantization of 
gauge theories. I had already reconciled myself to the
idea of spending the next few years there, when quite unexpectedly  
I was  informed that I could finally join CERN because another fellowship 
candidate had declined the offer from CERN. So we moved
to Geneva in November 1979. In the following years I continued 
to meet Julius regularly, rejoining the Karlsruhe institute as an associate
professor for the two years 1986 - 1988. After that contacts became less
frequent, as Julius accepted a joint offer from LMU Munich
and the MPI f\"ur Physik, and  finally moved there in 1990. 
Physicswise, he had quit supersymmetry activities to start 
exploring new ground, turning his full attention to 
non-commutative geometry where, as he told me more than
once, he saw a new El Dorado for theoretical physics.

At CERN my main focus switched to supergravity, an interest 
that had been ignited during a remarkable mathematical physics
(M$\cap$P) conference in Lausanne in 1978 that I was able to attend.
It was above all the brilliant performance of the French team represented by 
Bernard Julia and Jo\"el Scherk that got me hooked. They both talked about 
their recent progress with maximal supergravity. Bernard reported on his discovery
with Eug\`ene Cremmer of the hidden exceptional $\Es$ symmetry of maximal 
$N=8$ supergravity, while Jo\"el gave a wildly funny presentation 
of their recent construction of $D=11$ supergravity,
drawing all kinds of parallels between the superworld and
Nietzschean  philosophy. This is when my `extremist view' on supersymmetry
slowly started shaping up: if it is to be supersymmetry, then either as much
of it as possible, or none at all!  For this reason I decided not 
to get  involved in the activities on $N=1$ supersymmetry
model building, both with regard to further theoretical developments such 
as developing the tensor calculus for supergravity matter couplings, or concurrent
efforts to apply the theory to the phenomenology of the SM and its extensions.  
I rather concentrated my efforts on learning extended supergravity
in the component (not superspace) formalism. In doing my first steps
I was greatly helped by Paul Townsend, whose subtle sense of humor and very
British understatement I have always admired, and Peter van Nieuwenhuizen,
one of the fathers of supergravity, who generously accepted me as 
the ignoramus newcomer into the collaboration.

Around that time I started working with Bernard de Wit, aiming
right away for maximal $N=8$ supergravity. Not only had Bernard
done important preparatory work, but he also entrusted me with a 
big pile of hand-written notes. I spent the following weeks 
assiduously working my way through these notes,
until I reached a stage where I could no longer see the forest 
for all the trees. Since our initial attempts at gauging the 
maximal $N=8$ theory thus failed badly, we turned to $N=5$ supergravity
as a kind of stopgap measure, a theory that by itself is only of limited 
interest (except perhaps as a testing ground for checking unexplained
cancellations and UV finiteness properties of extended supergravities
beyond four  loops, as I learnt very recently from Zvi Bern).
There we finally got things to work! Having convinced ourselves
that $N=8$ would still refuse to cede, we decided to include an appendix 
in our $N=5$ paper explaining in all technical detail why $N=8$ supergravity 
could not be gauged. The paper was practically finished, so we decided to
follow our usual habit of having another cup of coffee in  the CERN
cafeteria before releasing the paper. And there it suddenly struck us
like a bolt of lightning out of the sky: we had made an elementary mistake 
in the calculation -- so trivial that I cannot even remember now what it was!
Needless to say that we rushed back to get to work right away on 
revising the paper. But then it still took several months to 
complete the construction of gauged $N=8$ supergravity

Optimism that $N=8$ supergravity might have something to do
with the real world was still running high in 1982:  on my first visit to DAMTP 
none less than Stephen Hawking assured me that, in his opinion,
this was the best candidate for a unified theory. He immediately
put a graduate student (Nick Warner) on the problem of finding
suitable `realistic' symmetry breaking vacua of the $N=8$ potential. 
Although I kept coming to Cambridge as a visitor numerous 
times afterwards, I could never quite figure out whether Stephen
continued to stand by this claim, or whether he had secretly switched
allegiance to string theory at some point. 
Independently, at around the same time there were attempts 
to relate the $N=8$ supergravity to `real physics', most notably 
by Ellis, Gaillard, and Zumino, who tried to  exploit a conjecture 
by Cremmer and Julia, according to which the (chiral!) SU(8) $R$-symmetry of the
theory could become dynamical. There was a short-lived burst of excitement when they
found three $\bar{\bf 5} \oplus {\bf 10}$ fermion multiplets of SU(5), but
those were accompanied by huge towers of `junk particles'  that were
obviously useless for realistic applications. Another, and 
to me even more intriguing, proposal was made by M. Gell-Mann,
and I will come back to it below. Furthermore, hopes for an `easy' 
argument that $N=8$ supergravity could be finite to all
orders were dashed when it was realized in a series of papers 
by Howe and Lindstr\"om, Kallosh and later Howe, Stelle and
Townsend that the theory admits (linearized) counterterms 
from three loops onward~\footnote{To be sure, the question of finiteness 
(or not) of $N=8$ supergravity remains up in the air,  in spite of stunning 
computational advances by Z. Bern, L. Dixon and collaborators, with results now
 reaching up to five loops. However, for all I know, the seven-loop calculation 
 that might finally settle the question remains beyond reach. {\em Idem} for
 the question whether the known `seed' counterterms can be made fully 
 compatible with non-linear supersymmetry and non-linear $\Es$ invariance.}.
So at that point it looked like all roads were blocked.

On top of these failings there was another major development that
pushed $N=8$ supergravity aside as a leading candidate theory for unification. This was 
the advent of heterotic string theory in 1984 which rolled over the CERN theory 
division like a tsunami, burying everything else underneath it. People
were busily scurrying to jump on the bandwaggon, 
being certain that all problems would soon be solved,
with an almost unique path from the heterotic theory to SM low energy physics.
The expectation that physics would be over within a few weeks 
resulted in a coherent state of collective inebriation that 
in retrospect looks rather absurd to me. Nevertheless, the general attitude from 
that moment on and for many years thereafter remained  that anyone 
{\em not} working on string theory and the heterotic string was completely 
out of the loop. Too bad for our poor colleagues who had invested all
their stock into non-string approaches to quantum gravity! 

In spite of these developments I decided to stick with maximally 
supersymmetric theories, and to persist with my decision 
not to join the booming MSSM and string model building industry 
(which at one point prompted John Ellis  to exclaim ``Hermann, we'll never turn 
you into  a good phenomenologist!"). Excursions into string theory 
proper were only sporadic, perhaps most notably with our proposal
(with A. Casher, F. Englert and A. Taormina) to derive all consistent
superstring theories from the $D=26$ bosonic string, including
a mechanism to get space-time fermions out of the bosons, and
thus explain the emergence of supersymmetry from a purely bosonic theory. 
Somewhat to my surprise, this proposal generated a huge amount of attention,
getting me an invitation from M. Gell-Mann to present these ideas at his 
newly founded Santa Fe Institute. There, however, I failed to convince
the fully assembled string establishment of the virtues of our insights. 
The main drawback was that despite all efforts our construction
remained entirely kinematical -- like most subsequent constructions 
of string vacua! -- with no compelling dynamical explanation why the
$D=26$ string should do such a thing. Towards the end of my stay at CERN
I also wrote one paper with Sergio Fubini, showing how to take
the `square root' of the Yang-Mills action in eight dimensions, in an
attempt to gain better
access to the maximal ($N=4$) super-Yang-Mills theory. This led to
the discovery of what we called the octonionic instanton.
Sergio, one of the true pioneers of string theory, was a very kind person.
He often told me what life was like at MIT when they did their 
ground-breaking work there, and hardly anyone would bother to show up at 
their seminars. Talk about volatility in the public perception of
what is good theory!

Over the following years I devoted much time to Kaluza-Klein supergravity 
and the question of consistent truncations beyond the linearized 
approximation, mostly in collaboration with B. de Wit (I think
the importance of this problem is still not fully appreciated
in work on the phenomenology of effective low energy actions obtained
from higher dimensions). Further topics were all kinds of things 
exceptional, such as quasi-conformal realizations and minimal unitary
representations of exceptional non-compact groups, and $\Ea$ 
in particular, where I learnt a lot from Murat G\"unaydin, 
and the construction of maximal gauged supergravity in $D=3$ with
Henning Samtleben, where again $\Ea$ played a pivotal role.
However, my greatest fascination 
remained with B. Julia's early papers conjecturing the appearance 
of the {\em infinite-dimensional} exceptional symmetries $\En$ and $\Ez$ in 
the further reduction to two and one dimensions. Of these, $\En$ symmetry
(or more precisely, E$_{9(9)}$) 
is now understood to be an extension of the Geroch group in general relativity,
but $\Ez$, the maximal rank hyperbolic Kac-Moody algebra,
remains a total enigma. What struck me most there was the unexpected link
between two seemingly disparate developments in physics and pure mathematics.
Namely, in both cases the search for distinguished structures leads to
unique answers. On the physics side, trying to make Einstein's theory 
as (super-)symmetric as possible, one arrives at a unique answer
($D=11$ maximal supergravity), while on the other side mathematicians
identified a uniquely distinguished Lie algebra, the maximally extended hyperbolic
Kac--Moody algebra $\Ez$, by way of classifying Dynkin diagrams.
In fact, $\Ez$ has been shown to contain all simply laced hyperbolic
Kac-Moody algebras as subalgebras, and is thus an all-encompassing 
mathematical entity -- much like $D=11$ supergravity, which keeps 
popping up in all discussions of M theory.

\section{Supersymmetry and fundamental physics}

The last time I saw Julius Wess, who had meanwhile moved to Hamburg,
was in Israel on the occasion of  Eliezer Rabinovici's and Shimon Yankielowicz' 
60th birthday celebrations in Jerusalem and Tel Aviv\footnote{This was the first
 time I broke ranks with the supersymmetry community when reporting about
 my work with K. Meissner, where we propose that it is {\em conformal symmetry}
 rather than supersymmetry that stabilizes the electroweak scale.}.
Julius seemed to be 
in great shape, and we agreed that he should soon visit me in Potsdam. 
So I was absolutely shocked when I learnt that he had suffered a stroke 
shortly after; he died in 2007, and the planned visit never came to pass.  
Tragically, he thus also missed the chance to see the outcome 
of the LHC experiment which he had been eagerly awaiting: over the years
he had become increasingly confident that LHC would find superparticles 
(whereas Bruno Zumino seemed more skeptical, as far as I could tell from 
his review talks that I was able to attend). Despite his optimism, and surely 
being aware that experimental confirmation of supersymmetry would earn them
a very nice reward, Julius was always modest  enough to admit the possibility 
that his expectations could turn out to be totally wrong. There was no question 
for him that experiment is the ultimate arbiter, and that either way one would
have to accept the outcome of the experiment. 

Half a century has now passed since the discovery of supersymmetry. During this time
the subject has developed enormously, with stupendous advances on
many fronts, some of which are also documented in this book. Supersymmetry 
has been a major driving force of  developments in mathematical physics and 
pure mathematics. So it is definitely here to stay!
Nevertheless, we now (in 2023) have to face up to the fact that supersymmetry, 
at least in the form championed over many years, is off the table 
as a realistic option for real life particle physics.
15 years of LHC searches have not produced a shred of evidence 
for superpartners of any kind.  Quite to the contrary,  the integrated results from LHC 
strongly indicate that the SM  could happily live up to the Planck scale more or less 
{\em as is}, and {\em without} supersymmetry or other major 
modifications.\footnote{One should also note the following point as possible
 evidence against supersymmetry.
 While supersymmetry strongly prefers BPS-type static configurations
 (such as AdS space), it is neither compatible with a positive cosmological 
 constant nor, more generally,  with a time-dependent cosmology -- but 
 this is what we see when we look out of the window!}

So where are we to go from here? Should we simply give up on
the idea that supersymmetry is of any relevance to the real world, as 
demanded by a growing chorus of voices? Or pin our hopes on the
next (100 TeV?) collider that, if it is ever built, will finally reveal that
Nature {\em is} supersymmetric? If such hopes rely on further 
pursuing the (no longer `low' energy) $N=1$ supersymmetry option
I am afraid they may well be disappointed, because 
the main motivating argument, namely solving the 
hierarchy problem, has largely evaporated. On the other hand, 
looking at the current disarray and confusion reigning in `Beyond the SM' 
particle physics, I feel that simply abandoning the idea of supersymmetry 
for `real' physics altogether would throw out the baby
with the bath. Without {\em some} guiding principle, 
we may simply end up in an unnavigable multiverse of ideas (just
look at the wildly diverse tableau of approaches to quantum gravity!), 
with no prospects of ever validating or falsifying any of them all the way.

To end this contribution on a more speculative note, let me outline
where I see a possible way forward.  This is to search for concepts 
{\em beyond} space-time supersymmetry, while taking the hints from what we have
learnt from supersymmetry studies over the years. In other words, throw 
out the bath, but try to save the baby! And here I would
first of all like to point out the remarkable fact
that, with the possible exception of the asymptotic safety program, 
and despite their differences, {\em all} current approaches to quantum gravity 
are united in the belief 
that something dramatic happens to space or space-time near the Planck scale, 
in the sense that classical space-time must dissolve or `de-emerge' there.
This is in analogy with continuum fluid dynamics which emerges from an 
underlying discrete structure of atoms and molecules. But if there is 
no space-time to begin with, the basic superalgebra
$$
\big\{ Q_\alpha^i \,,\, \bar{Q}_{{\dot\beta}j} \big\}
\,=\, 2 \delta^i_j \sigma^\mu_{\alpha\dot\beta} P_\mu
$$
and possibly even the distinction between space-time bosons and fermions,
cannot be fundamental. If consequently space-time, and concomitant
concepts such as general covariance  and gauge symmetries
are to be emergent,  the one million (or better: one billion?) dollar question is: 
{\em What}  do they emerge {\em from}? 

While there are no answers, there is also no lack of ideas and suggestions.
One of them is a specific proposal that I have been involved in
with Thibault Damour, Marc Henneaux and Axel Kleinschmidt, 
and which proceeds from the well-known BKL
(short for: Belinski-Khalatnikov-Lifshitz) analysis of cosmological
solutions of Einstein's equations near a space-like (cosmological) singularity.
According to this analysis spatial points decouple near the singularity,
effectively realizing a dimensional reduction to one
(time) dimension at the singularity.  But  
this is exactly the reduction for which one expects the hidden symmetry 
of maximal supergravity to get enlarged to the maximally extended 
hyperbolic Kac--Moody symmetry $\Ez$! The hypothesis is therefore 
that the true symmetry reveals itself only in a `near singularity limit', in perfect
analogy with the full electroweak symmetry becoming manifest only 
in the high energy limit. So $\Ez$, not supersymmetry, would be the key 
player!  Towards a more concrete realization of this idea we have 
proposed a `space-less' one-dimensional sigma model, mapping the
supergravity dynamics in space-time to a 
null-geodesic motion on the infinite-dimensional coset manifold $\Ez/K(\Ez)$.
While capturing some essential features of maximal supergravity
this ansatz remains incomplete for many reasons. The outstanding 
challenge is to explain in full detail how $\Ez$ can replace
space-time based QFT and the emergence of classical concepts such 
as space-time and gauge symmetries from a purely algebraic
construct.\footnote{P. West has proposed a conceptually very different 
 scheme based on the indefinite (but non-hyperbolic) Kac--Moody 
 algebra $E_{11}$. His scheme does not invoke dimensional reduction. }
One would expect quantum theory to play an essential role in
understanding how this happens. A further difficulty is
that we have hardly any idea how to physically interpret
the states associated with the imaginary roots of $\Ez$, starting out
with the `dual graviton'. One encouraging feature, especially in
connection with the incompatibility of supersymmetry with
time dependent cosmological solutions, is that the $\Ez/K(\Ez)$ 
sigma model does {\em not} admit static solutions.

What is clear, however, is that $\Ez$ goes beyond supersymmetry in the sense 
that it `knows' everything that supersymmetry `knows'. For instance, the information
about the bosonic constituents of the maximal supergravity multiplets
($D=11$, IIA, IIB,...) can be directly read off from the $\Ez$ Dynkin diagram,
simply by `slicing' it in different ways w.r.t. its
various finite subdiagrams. Likewise, the fermionic sectors of the 
maximal supergravity multiplets are governed by the maximal compact
(or `involutory') subgroup $K(\Ez)$ of $\Ez$, an infinite prolongation of 
the $R$-symmetry groups of maximal supergravities. For instance,
the IIA and IIB fermion multiplets at a given spatial point are again obtained 
by decomposing the $K(\Ez)$ under its finite-dimensional $R$-symmetry subgroups.
Moreover, one can expect that due to the unique properties of its
root lattice (being the unique even self-dual Lorentzian lattice 
in ten dimensions), $\Ez$ will eventually play a key role 
in ensuring the (non-perturbative) quantum consistency of the theory, 
in the same way that the perturbative finiteness and consistency of string theory
are ensured by {\em modular invariance}, whereas supersymmetry is mainly needed
to get rid of tachyons. So there is some evidence that, at the most basic level,
duality and modular symmetries are indeed more important than supersymmetry.

There is yet another, and different, argument pointing in the same direction,
though perhaps more obliquely. In 1983, M. Gell-Mann proposed what 
he described as ``a last ditch attempt to salvage the haplon interpretation
of $N=8$ supergravity". The SO(8) gauge group of the theory
contains a vector-like SU(3) $\!\times\!$ U(1), but it is immediately
evident that this SU(3) cannot be identified with the color symmetry SU(3)$_c$.
Noticing that after the removal of eight Goldstinos
(to break all supersymmetries) one is left with the right number $48 = 3 \times 16$ 
of spin-$\frac12$ fermions (corresponding to three generations
of quarks and leptons, including right-chiral neutrinos),
he had the very strange idea that the supergravity SU(3) should
be identified with the diagonal subgroup of SU(3)$_c$ and a hypothetical
family symmetry SU(3)$_f$. Putting the quarks and leptons into the 
appropriate representations of SU(3)$_c$ and SU(3)$_f$ 
(there is only one way to do this properly)
he obtained exact agreement! In addition,
the electric charge assignments almost work, in that the U(1) charges 
are systematically off by a `spurion charge' shift of $\pm \frac16$. So the proposal 
fails only by very little, and in a very systematic fashion. Not long after,
Nick Warner and I were able to show that this scheme is {\em dynamically} 
realized at one of the stationary points of the $N=8$ potential. For a few days 
we were ecstatic, trying very hard to push the agreement further  --
{\em alas}, in vain. Since the proposal was greeted with understandable skepticism 
by many colleagues (who had mostly placed their bets with $N=1$ low energy
supersymmetry), I eventually decided to put it aside, since I did not want to 
get trapped in a quixotic chase after a mirage.

Nevertheless, the idea stuck in my mind. After all,  if there are really only 
48 fundamental spin-$\frac12$ fermions,  no more, no less, what other scheme 
could possibly explain this, with a {\em unique} answer as the outcome?
Besides, the scheme can be immediately and easily falsified if only one
extra fundamental spin-$\frac12$ fermion were to be detected. In 2015, when it began
to dawn on everyone that LHC would not produce the predicted cornucopia of 
new particles, Krzysztof Meissner and I therefore decided to have another look,
by asking ourselves what it would take to rectify the mismatch 
of electric charges.  As it turned out, this could be 
achieved by a surprisingly simple deformation of the U(1) group by the generator
$$
\cI \,:=\, \frac{1}{2} \Big(T \wedge \id \wedge \id + 
\id \wedge \, T \wedge \id + \id \wedge \id \wedge \,T + T \wedge T \wedge T \Big) 
$$
acting on the tri-spinor $\chi^{ijk}$ of $N=8$ supergravity which transforms in the 
$\mathbf{56} \equiv \mathbf{8}\wedge\mathbf{8}\wedge\mathbf{8}$  of SU(8). 
Here, the SO(8) matrix $T$    
represents the imaginary unit in the SU(3) $\!\times\!$ U(1) breaking, 
with $T^2=-\id$, which in  turn implies $\cI^2=-\id$. Because the triple 
wedge product $T \wedge T \wedge T$ is \emph{not} an SU(8) element, 
there is no way of incorporating this  deformation into $N\!=\!8$ supergravity. 
The `spurion shift' associated with $\cI$ and needed for matching theory
and observation is therefore incompatible with supersymmetry.

But now the remarkable fact (demonstrated jointly with Axel
Kleinschmidt) is that, at least in the one-dimensional reduction, this
deformation {\em can} be incorporated into $K(\Ez)$. This could mean that 
matching the observed SM fermion spectrum with fundamental theory
requires more than merely picking the right Calabi-Yau manifold, 
GUT group or fermion multiplets, but may involve understanding in detail
how such structures could emerge from a space-time-less
pre-geometric context, and that we may have to go all the way to fancy 
infinite-dimensional symmetries such as $\Ez$ and $K(\Ez)$
to make things work. Then the question is 
no longer whether $N=8$ supergravity is the right theory.
Rather, it is whether and how $\Ez$ can replace space-time 
supersymmetry as a guiding principle, and what kind of
pre-geometric theory it is that lies beyond $N=8$ supergravity 
and realizes these symmetries. And, most importantly, whether 
and how the partial matching with SM physics can be completed
beyond merely kinematical considerations.

Since it could easily take another 50 years to figure all this out
(assuming there is any truth in these admittedly unorthodox ideas), 
Krzysztof Meissner and I have recently started looking for possible 
experimental signatures that might provide observational tests of our
hypotheses, in an attempt to search for the proverbial
needle in the haystack. This we propose to do 
by concentrating on the massive gravitinos 
which are the only other fundamental fermions that would emerge from 
the $N=8$ supermultiplet. Unlike the MSSM gravitino, these gravitinos do carry 
SM charges and would thus participate in (non-chiral) SM interactions, which 
makes them directly detectable at least in principle. Taking into account the U(1) shift 
also for the gravitinos (which again requires $K(\Ez)\,$) 
we have a new (stable) dark matter candidate here, 
but of a very unusual type. The gravitinos would have to be fractionally 
charged, extremely massive, and at the same time extremely rare 
(with an abundance of roughly one gravitino  in a cube of $\cO(10\,km)$ side length), 
so as to have escaped detection until now. We have argued that, if such
gravitinos were really found, we would be able to address two outstanding open 
problems in astrophysics, namely (1) explaining the origin of supermassive 
primordial black holes in the very early universe (via the condensation and gravitational
collapse of lumps of supermassive gravitinos in the early radiation period), and
(2) understanding the origin of the ultra-high energy cosmic ray events that have been 
observed over many years at the Pierre Auger Observatory in Argentina 
(via the mutual annihilation of supermassive gravitinos in the 
`skin' of neutron stars). We are currently exploring ways to search for such 
particles in upcoming underground experiments, such as JUNO, as such
objects are evidently far beyond the reach of high energy colliders.

\section{Conclusions}

Independently of whether the ideas sketched above are on the right track 
or not, I remain attached to the Einsteinian point of view that we should try to 
understand and explain first of all {\em our} universe and {\em our} 
low energy world, and that in the end there should emerge a more or less unique answer.
I believe that 50 years of supersymmetry have brought us a wee bit closer to 
this goal, though not as close as many would have wished. Of course, 
this point of view runs counter to currently prevalent views
according to which the only way out of the vacuum conundrum
of string theory is the multiverse. But if Nature must pick the `right' 
answer at random from a huge ($> 10^{272\,000}\,$?) number of 
possibilities, I see no hope that we would ever be able to confirm or 
refute such a theory, nor to reach a truly fundamental understanding
of the world that we see.

Already in 1929, and in connection with his first attempts at unification,
Albert Einstein published an article in which he states with wonderful 
and characteristic lucidity what the criteria should be of a `good' unified theory: 
(1) to describe as far as possible all phenomena and their inherent links, and 
(2) to do so on the basis of  a minimal number of assumptions and logically 
independent basic concepts. The second of these goals, also known as the 
principle of Occam's razor, he refers to as ``logical unity" (``logische Einheitlichkeit"), 
and goes on to say:  ``Roughly but truthfully, one might say: we not only want 
to understand {\em how} Nature works, but we are also after the perhaps utopian 
and presumptuous goal of understanding {\em why} Nature is the way it is, 
and not otherwise."\\[1mm]
To which I have nothing more to add!

\vspace{0.5cm}
\noindent
{\bf Acknowledgments:} I would like to thank Misha Shifman for giving me this
opportunity to provide my personal perspective on the story of
supersymmetry  which owes so much to the legacy of Julius Wess and Bruno Zumino.
I am also grateful to all my collaborators (including those, whose
names do not appear in this text) for the collaborations and lasting
friendships that I have been able to build over many years. My special
thanks go to Axel Kleinschmidt and Stefan Theisen for their
friendship and unwavering support over many years.

\end{document}